\documentclass[prl,onecolumn]{revtex4}
\usepackage{amsmath}
\usepackage{amssymb}
\usepackage{graphicx}
\usepackage{subfigure}
\usepackage{verbatim}
\usepackage{hyperref}
\usepackage{doi}
\usepackage{amsfonts}
\usepackage{bbm}
\usepackage[normalem]{ulem}
\usepackage{color}
\usepackage{datetime}
\longdate

\begin{document}
\title{A Bj\"orling Representation for Jacobi Fields on Minimal Surfaces and Soap Film Instabilities}
\author{Gareth P. Alexander}
\email{G.P.Alexander@warwick.ac.uk}
\affiliation{Department of Physics and Centre for Complexity Science, University of Warwick, Coventry, CV4 7AL, United Kingdom.}
\author{Thomas Machon}
\email{t.machon@bristol.ac.uk}
\affiliation{H. H. Wills Physics Laboratory, University of Bristol, Bristol BS8 1TL, United Kingdom.}
\begin{abstract}
We develop a general framework for the description of instabilities on soap films using the Bj\"orling representation of minimal surfaces. The construction is naturally geometric and the instability has the interpretation as being specified by its amplitude and transverse gradient along any curve lying in the minimal surface. When the amplitude vanishes, the curve forms part of the boundary to a critically stable domain, while when the gradient vanishes the Jacobi field is maximal along the curve. In the latter case, we show that the Jacobi field is maximally localised if its amplitude is taken to be the lowest eigenfunction of a one-dimensional Schr\"odinger operator. We present examples for the helicoid, catenoid, circular helicoids and planar Enneper minimal surfaces, and emphasise that the geometric nature of the Bj\"orling representation allows direct connection with instabilities observed in soap films. 
\end{abstract}
\date{\today}
\maketitle

\section{Introduction}
\label{sec:intro}

Minimal surfaces arise in many areas of science, including in lipid bilayers and bicontinuous phases~\cite{longley83,hyde84}, smectic liquid crystals~\cite{kamien99,matsumoto11}, membranes~\cite{andersson88,powers02}, the structural colour of butterfly wings~\cite{saranathan10}, liquid bridges~\cite{cryer92,chen97} and soap films~\cite{courant40}. Generally they occur in physical systems dominated by surface tension or similar geometric energies, which have minimal surfaces as critical points. A basic observation, studied readily in soap film realisations of minimal surfaces, is that they become unstable as the boundary is varied. The classic example of this is the instability of the catenoid, a problem studied by Plateau~\cite{Plateau}. For two circular rings of equal size, the instability occurs at a critical value of the ratio of the diameter of the rings to their separation (of about $1.509$) and leads to the rapid collapse of the soap film via `pinching' of the neck of the catenoid. For greater separations the only stable soap film is a pair of discs, one spanning each ring. Thus soap films, and minimal surfaces, provide a simple setting in which instabilities and singularity formation can be studied, including experimentally. More generally, these instabilities illustrate how surface tension serves to influence morphology, and drive morphological changes, in a simple but generic system. 

Recently, the nature of soap film instabilities and transitions has been considered for a variety of other surfaces~\cite{boudaoud99,goldstein10,goldstein12,goldstein14,pesci15,moffatt16,machon16}. For a soap film forming a non-orientable M\"obius strip, deformation of the boundary wire leads to an instability where the `neck' of the M\"obius strip collapses, as in the catenoid, but the resulting singularity occurs at a point of the boundary frame, rather than in the interior of the film, and the subsequent topological transition converts the non-orientable surface into a single, orientable disc~\cite{goldstein10}. The extension of this example to a family of similar soap films suggests that the singularities created in soap film collapses are always of one of these two types: the singularity occurs in the bulk, like the catenoid, or at the boundary, like the M\"obius strip~\cite{goldstein14,moffatt16}. Instabilities that do not lead to singularities are also possible and the prototype in this case is the helicoid~\cite{boudaoud99}. At a critical value of the ratio of the diameter of the helicoid to its pitch it becomes unstable and buckles into a ribbon. This instability can also be realised on a circular helicoid~\cite{machon16}, where if there are an odd number of half-turns the surface is non-orientable and the buckled state contains a localised `soliton' where the surface remains locally helicoidal. 

In addition to the aspect ratio of the frame (or analogous quantifier of the boundary curve), soap film instabilities may be analysed and characterised by determining the mode that becomes unstable. This is a mode of vibration of the soap film with zero frequency that is called in the mathematics literature a Jacobi field~\cite{barbosa1976,barbosa1980,meeks11}. For the helicoid (and catenoid) the Jacobi field corresponding to the observed instability has been found analytically~\cite{boudaoud99}, while for the Meeks minimal M\"obius strip it has been found numerically~\cite{pesci15}. In each case the Jacobi field is identified by solving for the normal modes of the minimal surface with a given boundary and varying the boundary, as is done in experiment, until the frequency of the lowest mode reduces to zero. A feature of the Jacobi fields for the catenoid and Meeks M\"obius strip is that their maximal amplitude coincides closely with the narrowest part of the neck in these surfaces. This has led to the general speculation that the systole, or curve of shortest length going around the neck, may be relevant in soap film collapses and singularities more generally~\cite{goldstein14,pesci15,moffatt16}. 

Here, we introduce a complementary approach in which an unstable mode is constructed around any curve on a minimal surface, using geometric methods that avoid directly solving the Jacobi equation. The bounding frame for a soap film that is critically stable with respect to this mode then follows from the nodal domain of the constructed Jacobi field. The approach is based on the Bj\"orling representation of minimal surfaces and constructs Jacobi fields through the solution of a one-dimensional differential equation (of Schr\"odinger type) defined along a curve. It therefore represents a dimensional reduction of the stability operator for the surface, leading to simplification in the analysis and description of the unstable modes when compared to previous methods~\cite{boudaoud99}. We present a number of explicit examples, some recapitulating known results and others providing new ones. This Bj\"{o}rling representation of Jacobi fields can be thought of as analytic continuation of a Jacobi field given data along a curve (the field and its transverse derivative), in the same way that the Bj\"{o}rling construction itself can be thought of as analytic continuation of a minimal surface given data on a curve.

The description of instabilities of minimal surfaces often focuses on a single mode of instability, for example that exhibited by a catenoid suspended between two circular rings of equal diameter. For rings of unequal diameter the conditions for producing instability vary, as does the unstable mode itself. There are then two questions: To what extent can the unstable mode be controlled or tuned? And, is there a `fundamental' or canonical instability for any given minimal surface? In response to the former question, we provide an explicit construction of an unstable mode localised around any curve that provides initial data for the Bj\"orling representation of the minimal surface, the nodal domain then specifies the boundary curve required for the instability to be realised on a soap film. For the latter, the suggestion of Goldstein {\sl et al.}~\cite{goldstein14,pesci15,moffatt16} that the systole is relevant provides one motivation for choosing a canonical curve, however, a full resolution for this question remains outstanding.

\section{Stability of Soap Films and Jacobi Fields}
\label{sec:stability}

The energetics of soap films are governed by surface tension and its static properties by its total surface area and the conformation of the bounding frame. The energy of a soap film $\Sigma$ is simply proportional to its total surface area  
\begin{equation}
E = \sigma \int_{\Sigma} dA ,
\end{equation}
where $\sigma$ is the surface tension, if we neglect the elasticity of the boundary wire~\cite{giomi12}. 
Critical points of the area functional are minimal surfaces and characterise the soap film morphology as a surface with vanishing mean curvature. Its instabilities are then found by studying the second variation of area, which leads to the Schr\"odinger-style stability operator $-\nabla^2+2K$, with Dirichlet boundary conditions, where $K$ is the Gaussian curvature of the surface and $\nabla^2$ its Laplace-Beltrami operator~\cite{simons68,Neel}. The eigenfunctions of the stability operator correspond to the normal modes of oscillation of the soap film, whose squared frequencies are given in terms of the eigenvalues, $\lambda_i$, by $\omega_i^2=\sigma \lambda_i/\rho h$ where $\rho$ and $h$ are the density and thickness of the soap film, respectively~\cite{boudaoud99}. The film is stable so long as all of the squared frequencies are positive; conversely, an instability is produced by manipulating the frame so as to push the lowest eigenvalue of the stability operator below zero. This happens generically~\cite{docarmo79,fischercolbrie80,pogorelov81,ros06}; heuristically, because the mean curvature is zero the Gaussian curvature is non-positive, and vanishes only for a disc, so that the eigenvalues are only positive if the eigenfunction has sufficiently large derivatives and as these can be made smaller through control of the boundary, instability is generic. 

The observed instabilities in soap films are associated to normal modes that are solutions of the Jacobi equation 
\begin{equation}
\bigl( - \nabla^2 + 2K \bigr) \psi = 0,
\label{eq:jacobi}
\end{equation}
known as Jacobi fields~\cite{meeks11}, and that are also groundstates for the stability operator and vanish on the bounding frame. They are produced experimentally through manipulation of the boundary shape, so determining this shape is key to characterising any instability. We observe that the shape of the boundary at critical stability coincides with a part of the zero level set of a Jacobi field. Conversely, the zeros of a Jacobi field define a nodal domain, which is critically stable since the interior of such a region possesses a positive Jacobi field by restriction. (It is known that a minimal surface is stable if it has a positive Jacobi field~\cite{meeks11}.) So the boundary of such a nodal domain corresponds to the conformation of a frame that supports a soap film on the verge of instability.

Jacobi fields have an equivalent characterisation to~\eqref{eq:jacobi} in terms of one-parameter families of minimal surfaces. Namely, if ${\bf X}_{\epsilon}$ is a one-paramter family of minimal surfaces, varying ${\bf X}\equiv{\bf X}_{0}$, then the normal deformation 
\begin{equation}
\psi = {\bf N} \cdot \biggl. \frac{d {\bf X}_{\epsilon}}{d\epsilon} \biggr|_{\epsilon = 0} ,
\label{eq:jacobi_variation}
\end{equation}
where ${\bf N}$ is the unit surface normal, is a Jacobi field~\cite{meeks11}. As a particular example, the family ${\bf X}_{\epsilon} = (1+\epsilon){\bf X}$ corresponds to simple scale transformations of the minimal surface and the associated Jacobi field $\psi = {\bf N}\cdot{\bf X}$ is called the support Jacobi field~\cite{meeks11}. For instance, for the catenoid we may take  
\begin{align}
& {\bf X}(u,v) = \bigl( \cosh v \cos u , \cosh v \sin u , v \bigr) , 
&& {\bf N}(u,v) = \bigl( \textrm{sech } v \cos u , \textrm{sech } v \sin u , -\tanh v \bigr) ,
\end{align}
and the support Jacobi field, $\psi = 1 - v \tanh v$, reproduces the well-known basic instability~\cite{boudaoud99}. Moreover, the unstable mode corresponds closely to the observation that the catenoid collapses by shrinking of the `neck'. We will use this description of Jacobi fields to give a general construction of soap film instabilities using one-parameter families of Bj\"orling data for the minimal surface.

\section{The Bj\"orling Representation of Minimal Surfaces}
\label{sec:bjorling}

There are two classical representations of minimal surfaces; the Weierstrass representation, and the Bj\"orling representation. The Weierstrass representation gives a minimal surface in terms of analytic data, namely a meromorphic function $g$ (Gauss map) and a holomorphic 1-form $dh$ (`height differential')~\cite{meeks11} 
\begin{equation}
{\bf X} = \textrm{Re} \int \biggl( \frac{1}{2} \biggl( \frac{1}{g} - g \biggr) , \frac{i}{2} \biggl( \frac{1}{g} + g \biggr) , 1 \biggr) \, dh .
\end{equation}
In contrast, the Bj\"orling representation corresponds to a solution of the initial value problem for a minimal surface~\cite{lopez18}. Although the construction is classical, we give our own short account as we will make free use of its basic properties in what follows. 

Given an embedded curve ${\bf c}(u)$ and a unit magnitude line element ${\bf n}(u)$, at each point normal to the curve, the integral 
\begin{equation}
{\bf X}(u,v) = \textrm{Re} \int^{z} \bigl[ {\bf c}^{\prime}(w) - i {\bf n}(w)\times {\bf c}^{\prime}(w) \bigr]\, dw ,
\label{eq:Bjorling}
\end{equation}
where $z=u+iv$ and ${\bf c}(z), {\bf n}(z)$ are holomorphic extensions of the initial data, defines a minimal surface in $\mathbb{R}^3$ that contains the given curve, ${\bf X}(u,0)={\bf c}(u)$, and whose normal ${\bf N}(u,v)$ along that curve is the given line field, ${\bf N}(u,0)={\bf n}(u)$. Simple examples include the catenoid, with Bj\"orling data ${\bf c}(u)={\bf n}(u)=\cos u \,{\bf e}_1 + \sin u \,{\bf e}_2$, the helicoid, with Bj\"orling data ${\bf c}(u)=u\,{\bf e}_3$ and ${\bf n}(u)=\cos u \,{\bf e}_1 + \sin u \,{\bf e}_2$, and the family of bent helicoids~\cite{meeks07,mira06}, with Bj\"orling data ${\bf c}(u)=\cos u \,{\bf e}_1 + \sin u \,{\bf e}_2$ and ${\bf n}(u) = \cos(pu/2) \bigl[ \cos u \,{\bf e}_1 + \sin u \,{\bf e}_2 \bigr] - \sin(pu/2) {\bf e}_3$, where $p$ is an arbitrary integer. Many other explicit examples have also been produced recently~\cite{lopez18}. 

The Bj\"orling construction parameterises the surface with conformal coordinates $z=u+iv$; the holomorphic tangent vector 
\begin{equation}
\partial_{u} {\bf X} - i \partial_{v} {\bf X} \equiv 2 \partial {\bf X} = {\bf c}^{\prime}(z) - i {\bf n}(z) \times {\bf c}^{\prime}(z) ,
\end{equation} 
is a holomorphic function, as well as a null vector, $\partial {\bf X} \cdot \partial {\bf X} = 0$. The metric has the form $ds^2=\Omega^2[du^2+dv^2]$, where the conformal factor, $\Omega$, is given by 
\begin{equation}
2\Omega^2 = 4 \overline{\partial} {\bf X} \cdot \partial {\bf X} = \overline{{\bf c}^{\prime}} \cdot {\bf c}^{\prime} + \overline{({\bf n}\times {\bf c}^{\prime})} \cdot ({\bf n}\times {\bf c}^{\prime}) - i \Bigl[ \overline{{\bf c}^{\prime}} \cdot ({\bf n}\times {\bf c}^{\prime}) - \overline{({\bf n}\times {\bf c}^{\prime})} \cdot {\bf c}^{\prime} \Bigr] .
\end{equation}
The surface normal, ${\bf N}(u,v)$, follows from 
\begin{equation}
i 4 \overline{\partial} {\bf X} \times \partial {\bf X} = 2 \Omega^2 {\bf N} ,
\end{equation}
while the second fundamental form, $II=e\, du^2 + f(dudv+dvdu) + g\, dv^2$ with $g=-e$ since the surface is minimal, is conveyed by the holomorphic function 
\begin{equation}
e-if = - 2 {\bf N} \cdot \partial \partial {\bf X} = 2 \partial {\bf X} \cdot \partial {\bf N} .
\end{equation} 
From this the Gaussian curvature follows as 
\begin{equation}
K = - \Omega^{-4} \bigl( e^2 + f^2 \bigr) = - \frac{|\overline{\partial}{\bf X}\times\partial{\bf X}\cdot\partial\partial{\bf X}|^2}{|\overline{\partial}{\bf X}\cdot\partial{\bf X}|^4} . 
\end{equation}
Finally, we mention that the Bj\"orling representation allows for the identification of Weierstrass data for the minimal surface. For any explicit case the integrand in \eqref{eq:Bjorling} can be put into the form 
\begin{equation}
\Bigl[ {\bf c}^{\prime}(w) - i {\bf n}(w)\times {\bf c}^{\prime}(w) \Bigr]\, dw = \begin{pmatrix} \frac{1}{2} \bigl( \frac{1}{g} - g \bigr) dh \\[1mm] \frac{i}{2} \bigl( \frac{1}{g} + g \bigr) dh \\[1mm] dh \end{pmatrix} ,
\end{equation}
from which the Gauss map $g$ and height differential $dh$ can be read off. 

There is enormous freedom in the choice of Bj\"orling data to represent any given minimal surface, ${\bf X}$. The curve ${\bf c}$ can be chosen to be any (inextendible) smooth curve lying in the surface; the Bj\"orling normal, ${\bf n}$, is then just the restriction of the surface normal, ${\bf N}$, to ${\bf c}$. In addition, the Bj\"{o}rling integral may be performed with respect to any parameterisation. So it is important to bear in mind that different choices of Bj\"orling data $({\bf c},{\bf n})$ do not necessarily correspond to different minimal surfaces; indeed for each minimal surface there is a vast equivalence class of Bj\"orling data. It is far from clear that there is a canonical choice of representative of this equivalence class, although we shall go part way to constructing such in our treatment of Jacobi fields. 
Locally, the equivalence class of Bj\"orling data for a fixed minimal surface are given to first order by 
\begin{align}
{\bf c}_{\epsilon}(u) & = {\bf X}(u,\epsilon \sigma(u)) = {\bf c}(u) + \epsilon \sigma(u) \, {\bf n} \times {\bf c}^{\prime} + O(\epsilon^2) , \label{eq:equiv_Bc} \\
{\bf n}_{\epsilon}(u) & = {\bf N}(u,\epsilon \sigma(u)) = {\bf n}(u) - \epsilon \sigma(u) \, {\bf n} \times {\bf n}^{\prime} + O(\epsilon^2) , \label{eq:equiv_Bn}
\end{align}
where $\sigma(u)$ is an arbitrary smooth function and $\epsilon$ is infinitesimal. At any point on the Bj\"{o}rling curve, we may write $|c^\prime|^2 {\bf n}^\prime = ({\bf n}^\prime \cdot {\bf c}^\prime) {\bf c}^\prime +  ({\bf n}^\prime \cdot {\bf n} \times {\bf c}^\prime) {\bf n} \times {\bf c}^\prime$, which gives the following expression
\begin{align}
{\bf n}_{\epsilon}(u) & = {\bf n} + \epsilon \frac{\sigma(u)}{|c^\prime|^2} \left( ({\bf c}^{\prime \prime} \cdot {\bf n}) \, {\bf n} \times {\bf c}^\prime - ({\bf n}^\prime \times {\bf c}^\prime \cdot {\bf n}) \, {\bf c}^\prime \right) + O(\epsilon^2) . 
\end{align}
If we choose an arc-length parameterisation of ${\bf c}$, then we find
\begin{align}
{\bf c}_{\epsilon}(u) & = {\bf X}(u,\epsilon \sigma(u)) = {\bf c}(u) + \epsilon \sigma(u) \, {\bf n} \times {\bf c}^{\prime} + O(\epsilon^2) , \\
{\bf n}_{\epsilon}(u) & = {\bf N}(u,\epsilon \sigma(u)) = {\bf n} + \epsilon \sigma(u) \left( \kappa_n \, {\bf n} \times {\bf c}^\prime - \tau_r \, {\bf c}^\prime \right) + O(\epsilon^2) ,  
\end{align}
where $\kappa_n$ and $\tau_r$ are the normal curvature and relative torision of the Darboux frame associated to the Bj\"{o}rling curve.

\section{Construction of Jacobi Fields}

The Jacobi fields that describe instabilities of soap films can be constructed from a one-parameter family of minimal surfaces ${\bf X}_{\epsilon}$, varying the surface ${\bf X}\equiv {\bf X}_{0}$. Locally, minimal surfaces can be described by the pair of data $({\bf c},{\bf n})$ entering the Bj\"orling representation, so it follows that we can construct Jacobi fields from a one-parameter variation of this Bj\"orling data. Specifically, if $({\bf c}_{\epsilon},{\bf n}_{\epsilon})$ is any one-parameter variation of Bj\"orling data for the surface ${\bf X}$, then 
\begin{equation}
\psi = {\bf N} \cdot \biggl. \textrm{Re}\, \frac{d\;}{d\epsilon} \int^{z} \bigl[ {\bf c}_{\epsilon}^{\prime}(w) - i {\bf n}_{\epsilon}(w) \times {\bf c}_{\epsilon}^{\prime}(w) \bigr] dw\; \biggr|_{\epsilon=0} , 
\label{eq:BJ1}
\end{equation}
defines a Jacobi field. 
Now, in any deformation of a surface, motions that are tangential to the surface do not constitute any change and are equivalent to reparameterisations, so that it suffices to consider normal motions. This extends to the deformation of the Bj\"orling curve; deformations in the direction of the tangent ${\bf c}^{\prime}$ are equivalent to reparameterisations. In addition, the component in the (curve normal) direction ${\bf n}\times{\bf c}^{\prime}$ does not lead to any change in the minimal surface by~\eqref{eq:equiv_Bc}. Therefore, the general form of the deformation of the Bj\"orling curve generating an actual deformation of the surface is 
\begin{equation}
{\bf c}_{\epsilon}(u) = {\bf c}(u) + \epsilon \phi(u) {\bf n}(u) ,
\label{eq:deform_Bc}
\end{equation}
where $\phi$ is any function, periodic if the surface is orientable, $\phi(u+2\pi)=\phi(u)$, and antiperiodic, $\phi(u+2\pi)=-\phi(u)$, if the surface is non-orientable. A general deformation of the Bj\"orling normal, which preserves the orthogonality condition ${\bf n}_{\epsilon}\cdot{\bf c}_{\epsilon}^{\prime}=0$ to first order in $\epsilon$, is  
\begin{equation}
{\bf n}_{\epsilon}(u) = {\bf n}(u) - \frac{\epsilon}{|{\bf c}^{\prime}(u)|^2} \Bigl[ \theta(u) {\bf n}(u)\times{\bf c}^{\prime}(u) + \phi^{\prime}(u) {\bf c}^{\prime}(u) \Bigr] + O(\epsilon^2) ,
\label{eq:deform_Bn}
\end{equation}
where $\theta(u)$ is an arbitrary periodic function. Together, \eqref{eq:deform_Bc} and \eqref{eq:deform_Bn} lead to the Jacobi field  
\begin{equation}
\psi = {\bf N} \cdot \textrm{Re} \biggl( \phi(z) {\bf n}(z) - i \int^{z} \bigl[ \phi\, {\bf n} \times {\bf n}^{\prime} + \theta\, {\bf n} \bigr] \, dw \biggr) .
\label{eq:BJ}
\end{equation}

\begin{figure}
\begin{center}
\includegraphics{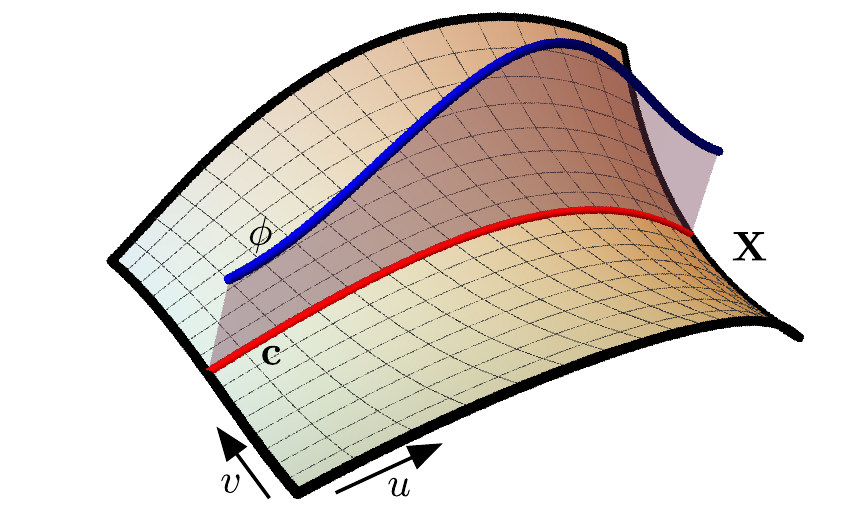}
\end{center}
\caption{Bj\"{o}rling representation of a Jacobi field. The minimal surface ${\bf X}$ contains the red Bj\"{o}rling curve ${\bf c}$, the line $v=0$. In the Bj\"{o}rling representation of a Jacobi field, \eqref{eq:BJ} the function $\phi$ gives the restriction of the Jacobi field to the Bj\"{o}rling curve, and is the surface deformation in the normal direction. The additional function $\theta$ gives the transverse derivative of the Jacobi field at $v=0$ and along with $\phi$ specifies the Jacobi field.}
\label{fig:diagram}
\end{figure}

For given Bj\"orling data, Jacobi fields are described by the pair of functions $(\phi,\theta)$. Evaluating \eqref{eq:BJ} on the curve ${\bf c}$ ($v=0$) gives that $\psi(u,0)=\phi(u)$, so that $\phi$ is the restriction of the Jacobi field to the Bj\"orling curve. A direct calculation also gives that $\partial_v\psi(u,0)=\theta(u)$, so that $\theta$ has the interpretation of being the transverse derivative of the Jacobi field along the Bj\"orling curve; if we think of the Jacobi field as originating from a 1-parameter family of minimal surfaces, then $\theta$ describes the local twisting of the surface along the Bj\"{o}rling curve. We remark that, for a fixed Jacobi field on a fixed minimal surface, there is an equivalence among the pairs of functions $(\phi,\theta)$ induced by the equivalence of Bj\"orling data, so that the choice of these functions is not independent of the choice of Bj\"orling data. A schematic of this construction of Jacobi fields is shown in Fig.~\ref{fig:diagram}. 

We highlight two choices as having a certain naturality. First, the choice $\phi=0$ corresponds to the Bj\"orling curve being (part of) the boundary of the critically stable domain. This choice is appealing because of how directly it connects to the experimental protocol of varying the bounding frame of a soap film until critical stability is reached. Second, the choice $\theta=0$ corresponds to the Jacobi field having its (transverse, local) maximum along the Bj\"orling curve. Although we will give examples of both choices we focus further formal development on the second. 

Continuing the construction of Jacobi fields with $\theta=0$, if we consider the Jacobi equation~\eqref{eq:jacobi}, multiplied by $\Omega^2$, and restrict it to the Bj\"orling curve ${\bf c}$ we find 
\begin{equation}
\Bigl. - \partial_{vv} \psi \Bigr|_{v=0} - \partial_{uu} \phi + 2 \Bigl. \Omega^2 K \Bigr|_{v=0} \phi = 0 .
\end{equation}
It follows that $\partial_{vv}\psi|_{v=0}$ is negative, and maximally so, if we choose $\phi$ to be the lowest eigenfunction of the stability operator restricted to the Bj\"orling curve, {\sl i.e.} of the operator $-\partial_{uu} + 2\Omega^2 K |_{v=0}$. This operator depends only on the Bj\"orling data and is explicitly given by 
\begin{equation}
- \partial_{uu} - 2 \frac{|{\bf n}\cdot{\bf c}^{\prime\prime}|^2 + |{\bf n}\cdot{\bf n}^{\prime}\times{\bf c}^{\prime}|^2}{|{\bf c}^{\prime}|^2} ,
\label{eq:dim_red}
\end{equation}
which can be rewritten in terms of the Darboux invariants, $\kappa_n$ and $\tau_r$, of the Bj\"{o}rling curve as
\begin{equation}
- \partial_{uu} - 2 \bigl( \kappa_n^2+\tau_r^2\bigr).
\label{eq:dim_red2}
\end{equation}
This leads to an appealing natural construction of Jacobi fields, given by \eqref{eq:BJ} with $\theta=0$ and $\phi$ the lowest eigenfunction of \eqref{eq:dim_red}. More precisely, we choose the lowest eigenfunction with periodic (bosonic) boundary conditions if the surface is orientable and the lowest eigenfunction with anti-periodic (fermionic) boundary conditions if the surface is non-orientable. The associated Jacobi field is then localised as strongly as possible around the curve ${\bf c}$, which corresponds to its maximum in the transverse direction. This analysis also amounts to a dimensional reduction of the original problem; rather than a two-dimensional Schr\"odinger equation we need only solve a one-dimensional version. Note that the choice of Bj\"orling data $({\bf c},{\bf n})$ remains arbitrary and the surface exhibits an instability for every possible choice. 

The Bj\"{o}rling construction, as well as our construction of Jacobi fields, becomes further geometrically special when the Bj\"{o}rling curve is chosen to be a geodesic; its tangent vector is parallel transported along itself and it follows that the surface normal is equal to the Frenet-Serret normal of the Bj\"orling curve. Recall that the Frenet-Serret frame of a curve ${\bf c}$ is the orthonormal triad $\{{\bf t},{\bf n},{\bf b}\}$ where ${\bf t}=d{\bf c}/ds$ is the unit tangent vector, with $s$ arc length, and 
\begin{align}
& \frac{d{\bf t}}{ds} = \kappa {\bf n} , 
&& \frac{d{\bf n}}{ds} = - \kappa {\bf t} + \tau {\bf b} , 
&& \frac{d{\bf b}}{ds} = - \tau {\bf n} .
\end{align}
Here, $\kappa$ is the curvature and $\tau$ the torsion of the curve ${\bf c}$. For geodesic Bj\"orling curves parameterised by arc length~\eqref{eq:dim_red2} becomes 
\begin{equation}
- \partial_{ss} - 2 \bigl( \kappa^2 + \tau^2 \bigr) .
\end{equation}

\section{Examples}
\label{sec:examples}

Here we give several examples illustrating how our construction can be used to give Jacobi fields for a variety of minimal surfaces. We emphasise that the natural geometric character of the Bj\"{o}rling representation allows one to easily reproduce instabilities observed in soap films. 

\subsection{Helicoid}
\label{subsec:helicoid}

\begin{figure}
\begin{center}
\includegraphics{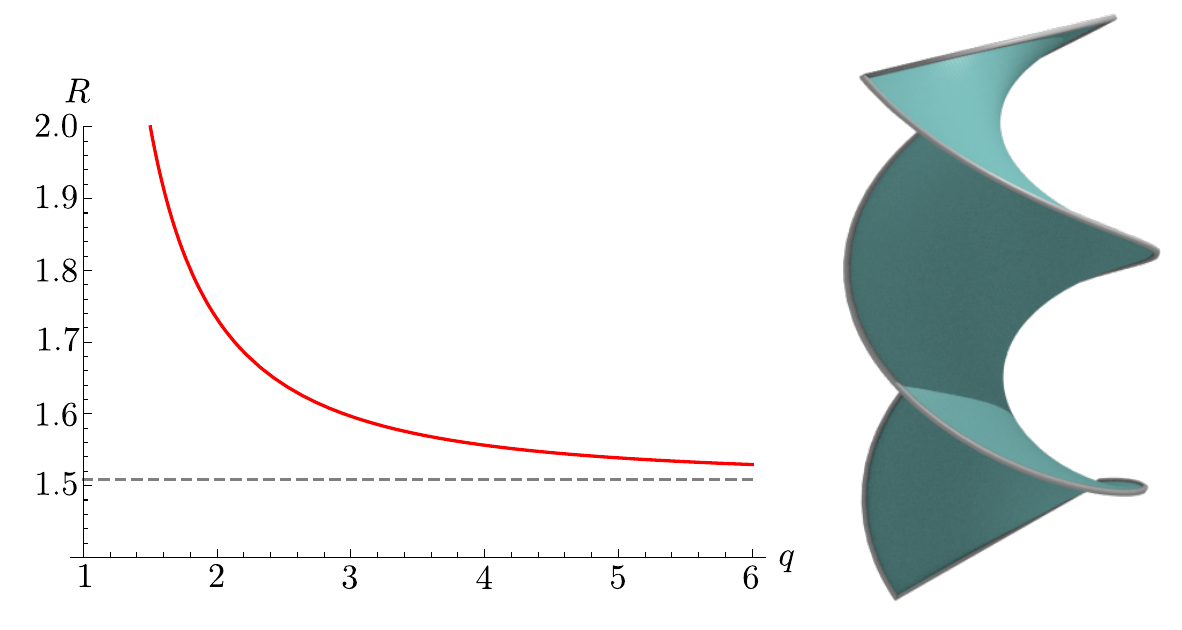}
\end{center}
\caption{Left: Critical radius $R$ as a function of number of half-twists, $q$, for a helicoid (red), with the asymptotic value (gray) determined by the root of \eqref{eq:BJhelicoid0}. Right: Critical helicoid with $q=2$.}
\label{fig:helicoid_figure}
\end{figure}

The helicoid is the surface ${\bf X}(u,v) = \sinh v \sin u \,{\bf e}_1 - \sinh v \cos u \,{\bf e}_2 + u{\bf e}_3$ and may be described by the Bj\"orling data ${\bf c} = u \,{\bf e}_3$, ${\bf n} = \cos u \,{\bf e}_1+\sin u \,{\bf e}_2$. With this choice \eqref{eq:dim_red} is $-(\partial_{uu}+2)$, whose eigenfunctions are $\cos ku$ and $\sin ku$ with eigenvalues $k^2-2$. The lowest eigenfunction corresponds to $k=0$ and gives $\phi=1$ and the associated Jacobi field \eqref{eq:BJ} recovers the well-known form~\cite{boudaoud99} 
\begin{equation}
\psi = 1 - v \tanh v .
\label{eq:BJhelicoid0}
\end{equation}
Choosing instead $\phi(u) = \cos(u/q)$ imposes that $\phi$, and hence the Jacobi field $\psi$, vanishes at $u=\pm q\pi/2$. This may be interpreted either as confining ends to the wire frame as in the experiments of Boudaoud {\sl et al.}~\cite{boudaoud99} or, if $q$ is an odd integer, periodic boundary conditions for a non-orientable strip~\cite{machon16}. In either case the associated Jacobi field is 
\begin{equation}
\psi = \cos(u/q) \Bigl[ \cosh(v/q) - q \sinh(v/q) \tanh v \Bigr] ,
\label{eq:BJhelicoid}
\end{equation}
which recovers the results of \cite{boudaoud99,machon16} and tends to~\eqref{eq:BJhelicoid0} as $q \to \infty$. Note that the nodal domain is only bounded in the $v$-direction if $q>1$; a portion of a helicoid containing less than (or exactly) one full half-twist is stable. We note in particular that this represents a substantial simplification of previous results~\cite{boudaoud99}, where the Jacobi fields were expressed using hypergeometric functions arising from a Schr\"{o}dinger equation with a modified P\"{o}schl-Teller potential. As an application, we can easily compute the critical radius for stability of a helicoid on a frame with $q$ twists, shown in Figure~\ref{fig:helicoid_figure}.

\subsection{Catenoid}
\label{subsec:catenoid}

\begin{figure}
\begin{center}
\includegraphics{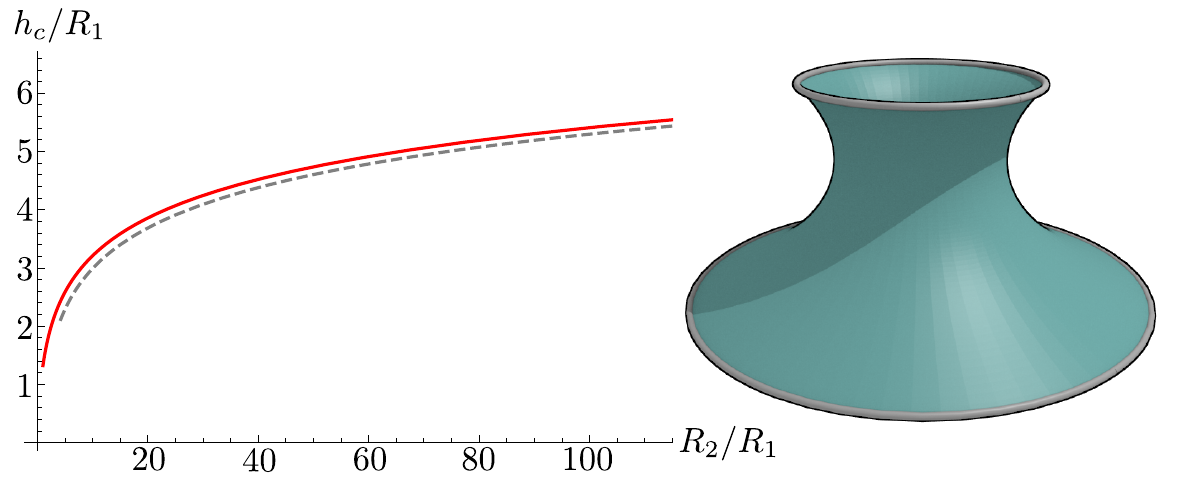}
\end{center}
\caption{Left: Critical separation $h_c/R_1$ as a function of the ratio of radii $R_2/R_1$ for catenoid (red) with asymptotic formula $\textrm{arccosh}(R_2/R_1)$ (gray). Right: Critical Catenoid with $R_2/R_1=2$.}
\label{fig:catenoid_figure}
\end{figure}

The catenoid and the helicoid are conjugate surfaces, connected through the Bonnet transform. They are isometric, from which it follows that they share the same Jacobi fields. Nonetheless, we also consider the catenoid explicitly, but focusing on different aspects. The catenoid is the surface ${\bf X} = \cosh v \cos u \,{\bf e}_1 + \cosh v \sin u \,{\bf e}_2 + v \,{\bf e}_3$ and may be represented by the Bj\"orling data 
\begin{align}
{\bf c}(u) & = \cosh v_0 \bigl[ \cos u \,{\bf e}_1 + \sin u \,{\bf e}_2 \bigr] + v_0 {\bf e}_3 , \\
{\bf n}(u) & = \textrm{sech } v_0 \bigl[ \cos u \,{\bf e}_1 + \sin u \,{\bf e}_2 \bigr] - \tanh v_0 {\bf e}_3 , 
\end{align}
for any constant $v_0$, amongst many other choices. Here, $v_0$ serves to provide a parameterisation of different choices for the Bj\"orling data allowing us to study how the associated Jacobi fields depend on this choice. The operator \eqref{eq:dim_red} is 
\begin{equation}
- \partial_{uu} - 2 \Bigl[ 1 - \tanh^4(v_0) \Bigr] , 
\end{equation}
whose lowest eigenfunction is simply the constant $\phi=1$ for any value of $v_0$ and the associated Jacobi field is 
\begin{equation}
\psi(v;v_0) = \textrm{sech}^2 v_0 \Bigl[ 1 - v \tanh(v+v_0) \Bigr] + \tanh v_0 \tanh(v+v_0) .
\end{equation}
The nodal set of this field determines the critical separation for the catenoid soap film spanning a pair of rings of unequal sizes. To leading order the solutions of $\psi=0$ can be estimated as $ -v_0$ and $1+\frac{1}{4} \exp(2 v_0)$ as $v_0 \to \infty$, so that for two rings of radii $R_2>R_1$, the critical separation grows as $\textrm{arccosh}(R_2/R_1)$, as shown in Figure \ref{fig:catenoid_figure}. 

As for the helicoid, we may also consider the eigenfunction $\phi=\cos(u/q)$, which leads in the case $v_0=0$ to the same Jacobi field \eqref{eq:BJhelicoid}. In this case the interpretation is that the minimal surface is only a `wedge' of the catenoid, $u\in[-q\pi/2,q\pi/2], q<2$. Again, the nodal domain is only bounded in the $v$-direction if $q>1$, meaning that half a catenoid (or less) is stable, an example of the half-space theorem~\cite{meeks11}. 

There is an alternative representation of the instabilities on the catenoid. Rather than taking the Bj\"{o}rling curve to be the maximum of $\psi$, we can take the Bj\"{o}rling curve to be a boundary, this implies we make the choice $\phi=0$, $\theta=\textrm{const}$, and yields the Jacobi field
\begin{equation}
\psi_\theta(v;v_0) = \theta \Bigl[ \tanh(v+v_0) \bigl( 1 + v \tanh v_0 \bigr) - \tanh v_0 \Bigr] . 
\label{eq:jacobi_catenoid_theta}
\end{equation}
By construction there is a node at $v=0$; for small $v_0 > 0$ there is also a node at $v \approx - (1 + v_0^{-1})$, which recovers our asymptotic estimate for the critical separation in terms of the ratio of radii. A beautiful description of the critically stable catenoid was given by Lindel\"of~\cite{lindelof1869} and can be readily verified using~\eqref{eq:jacobi_catenoid_theta}. Lindel\"of's construction says that lines tangent to the critically stable surface drawn from the two boundary rings, along the catenary direction, meet at a point on the symmetry axis. Such lines have tangent vector 
\begin{equation}
{\bf t} = \tanh(v+v_0) \bigl[ \cos u \,{\bf e}_1 + \sin u \,{\bf e}_2 \bigr] + \textrm{sech}(v+v_0) \,{\bf e}_3 ,
\end{equation}
and meet the symmetry axis at the point $\bigl(0,0,v+v_0 - \textrm{coth}(v+v_0)\bigr)$. Using that one of the boundaries is $v=0$, Lindel\"of's construction identifies the other as a solution of 
\begin{equation}
\textrm{coth}(v+v_0) - v = \textrm{coth } v_0 ,
\end{equation}
and routine manipulation shows that this is equivalent to the vanishing of the Jacobi field~\eqref{eq:jacobi_catenoid_theta}. 

In general there are three free parameters for this family of Jacobi fields, $v_0$, $\phi$ and $\theta$, with one redundancy. Defining the new coordinate $x = v+v_0$, measuring distance along the symmetry axis from the waist, this family is given by 
\begin{equation}
\begin{split}
\psi(x) & = \tanh v_0 \tanh x \bigl( \theta (x-v_0) + \phi \bigr) + \theta \bigl( \tanh x - \tanh v_0 \bigr) \\
& \quad + \phi \;\text{sech}^2 v_0 \bigl[ (v_0-x) \tanh x + 1 \bigr] .
\end{split}
\end{equation}
Then the Jacobi field is specified by its value and derivative at $x=0$ (for instance),
\begin{gather}
\psi|_{x=0} = \phi \; \textrm{sech}^2v_0 - \theta \tanh v_0,\\
\bigl. \partial_{x} \psi \bigr|_{x=0}  = v_0 \bigl( \phi \; \textrm{sech}^2v_0 - \theta \tanh v_0 \bigr) + \theta + \phi \tanh v_0 .
\end{gather}
Varying $v_0$, $\phi$ and $\theta$ while these two quantities (and hence the Jacobi field) are held fixed can be thought of as a nonlinear gauge transformation.

\subsection{Planar Enneper Surfaces}
\label{subsec:enneper}

We now turn to an example where the Bj\"{o}rling curve is a closed geodesic. The potential term in the one-dimensional Schr\"odinger operator is given by the Frenet-Serret invariants as $-2(\kappa^2+\tau^2)$. The eigenfunctions are therefore particularly simple if we choose a space curve for which $\kappa^2+\tau^2$ is constant, and in this case we may write
\begin{equation}
\kappa= \cos\bigl( \theta(s) \bigr), \quad \tau = \sin\bigl( \theta(s) \bigr) ,
\end{equation}
where $\theta(s)$ is a function of arclength, $s$. The simplest choice is to take $\theta(s)$ constant, which gives circles and helices as curves, and the helicoid/catenoid family of minimal surfaces. The next simplest choice is $\theta(s) = \alpha s$, with $\alpha$ constant. In this case the tangent, normal and binormal satisfy the equations
\begin{equation}
\frac{d}{ds}\begin{pmatrix}
\cos \alpha s \,{\bf t}+ \sin \alpha s \,{\bf b} \\
{\bf n}\\
\cos \alpha s \,{\bf b}- \sin \alpha s \,{\bf t}
\end{pmatrix}= \begin{pmatrix}
0 & 1 & -\alpha \\
-1 & 0 & 0 \\
\alpha & 0 & 0
\end{pmatrix} \cdot \begin{pmatrix}
\cos \alpha s \,{\bf t}+ \sin \alpha s \,{\bf b} \\
{\bf n}\\
\cos \alpha s \,{\bf b}- \sin \alpha s \,{\bf t}
\end{pmatrix}.
\end{equation}
Solving this gives the Bj\"{o}rling curve explicitly. For the curve to be closed we find
\begin{equation}
\frac{\alpha}{\sqrt{1+\alpha^2}} = \frac{p}{q} ,
\end{equation}
where $|p|<|q|$ are coprime integers, giving a total length of $2 \pi \sqrt{q^2-p^2}$. We then find that
\begin{equation}
{\bf c} = \begin{pmatrix}
\frac{2 p q \cos q r \sin p r - (p^2+q^2) \cos pr \sin q r}{q \sqrt{q^2-p^2}} \\
\frac{ (p^2+q^2) \cos pr \cos q r+ 2 p q \sin q r \sin p r }{q \sqrt{q^2-p^2}}\\
\frac{(q^2-p^2) \cos p r}{p q}
\label{eq:pecurve}
\end{pmatrix} ,
\end{equation}
where $r = s/\sqrt{q^2-p^2}$. This gives a family of closed space curves, indexed by integers $(p,q)$ with $\kappa^2+\tau^2=1$, it is easy to see that they are only embedded for $p=\pm(|q|-1)$, and we note that the curves live on a hyperboloid, satisfying the equation
\begin{equation}
\frac{1}{4} \left (\frac{q^2}{p^2}-1 \right) (c_1^2+c_2^2) - \frac{1}{4} c_3^2 = 1.
\end{equation}
As ${\bf c}$ is a geodesic the Bj\"{o}rling data is completely specified and surprisingly leads to a family of known minimal surfaces, the planar Enneper surfaces, and as a corollary we have shown that these surfaces have a closed geodesic with constant Gaussian curvature. One can contrast the curves we use \eqref{eq:pecurve} with the alternate choice of non-geodesic space curves based on torus knots, given by L\'{o}pez and Weber~\cite{lopez18}. We note that one may also choose the non-embedded case $p\neq\pm(|q|-1)$, which generates a family of immersed minimal surfaces. 

\begin{figure}
\begin{center}
\includegraphics{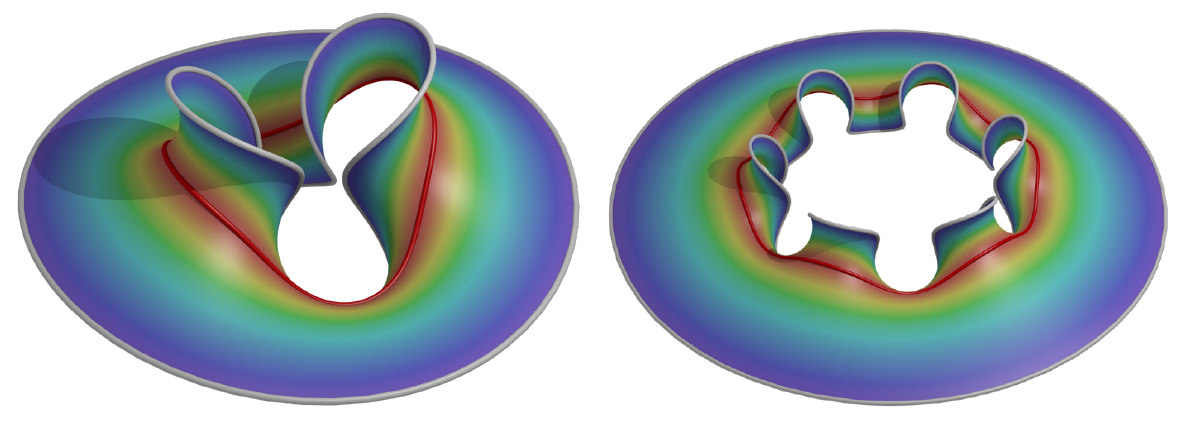}
\end{center}
\caption{Critical domains of planar Enneper surfaces and associated Jacobi fields. The red line indicates the closed geodesic along which the Gaussian curvature equals $-1$. The curvature and torsion of the geodesic are equal to $\kappa = \cos \alpha s$ and $\tau = \sin \alpha s$ respectively, where $s$ is arclength and $\alpha/\sqrt{1+\alpha^2}=p/q$. The hue indicates the strength of the Jacobi field, which attains a maximum value of 1 along the central geodesic and is equal to zero along the two boundary components. The left surface corresponds to $q=3,p=2$ surface, right $q=7, p=6$.}
\label{fig:planar_enneper}
\end{figure}

By construction, the Bj\"{o}rling-Jacobi operator along the geodesic is particularly simple and its eigenfunctions are (co)sinusoidal waves, so we take
\begin{equation}
\phi(s) = \cos \biggl( \frac{n s}{\sqrt{q^2-p^2}} + \chi \biggr) ,
\end{equation}
where $n \in \{0,1,2,\ldots\}$ and $\chi$ is a phase. Defining a rescaled variable $s^\prime = s/\sqrt{q^2-p^2}=u+i w$ and taking $p=q-1$, the Jacobi field associated with the eigenmode $\phi = \cos mu$ is 
\begin{equation}
\begin{split}
\psi_{q}^{m}(u,v) & = \frac{2q-1}{q(2q-1+\mathrm{e}^{2qv})} \Biggl\{ \Bigl( 2q - 1 - \mathrm{e}^{2qv} \Bigr) \biggl[ \frac{q-1}{2q-1} \cosh mv + \frac{1}{m} \sinh mv \biggr] \\
& \quad + 2 \cosh mv + \mathrm{e}^{qv} \biggl[ \frac{m+1}{q+m} \sinh (q+m)v - \frac{m-1}{q-m} \sinh (q-m)v \biggr] \Biggr\} \cos mu .
\end{split}
\label{eq:enneper_psi_m}
\end{equation}
Finally, choosing $m=0$ we can compute a critical domain for the planar Enneper surfaces, shown in Figure~\ref{fig:planar_enneper}.

\subsection{Circular Helicoids}
\label{subsec:circular_helicoid}

\begin{figure}
\begin{center}
\includegraphics{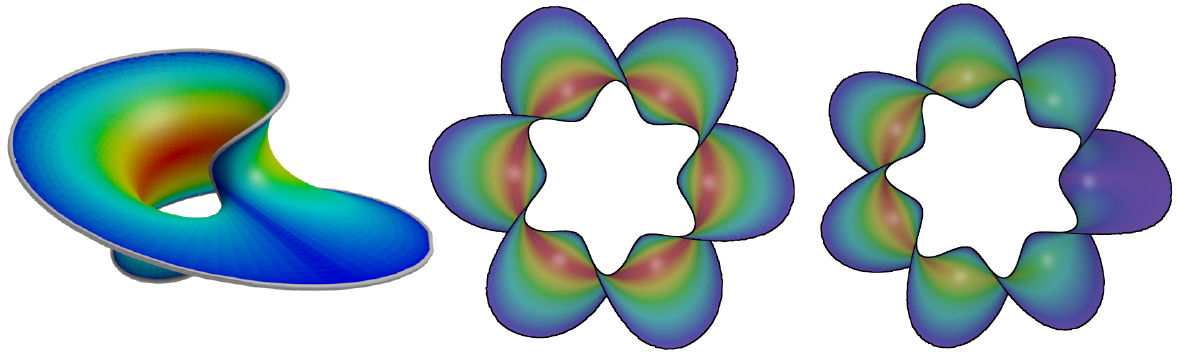}
\end{center}
\caption{Critical domains of twisted helicoids and associated Jacobi fields. Left: critical domain of Meeks' M\"{o}bius strip computed using the support Jacobi field, alternatively through the Bj\"{o}rling construction using the choices $\phi = \sin (u/2)$, $\theta = \frac{1}{2} \sin u$. The hue indicates the absolute value of the the Jacobi field. Centre: twisted strip with 6 half-twists, with a $(6,2)$ torus link boundary. The hue indicates the strength of the Jacobi field, equal to $\textrm{ce}_0\bigl( pu/2,2/p^2 \bigr)$ along the central line and zero on the boundary. Right: twisted strip with 7 half-twists, with a $(7,2)$ torus knot boundary. The hue indicates the absolute value of the the Jacobi field, equal to $\textrm{se}_{1/p}\bigl( pu/2,2/p^2 \bigr)$ along the central line and zero on the boundary. Because the surface is nonorientable, the Jacobi field has a nodal line which is clearly visible on the right of the surface.}
\label{fig:fig_twist}
\end{figure}

Our final example are the circular helicoids. We take these to be the surfaces 
\begin{equation}
\begin{split}
{\bf X}(u,v) & = \cosh v \Bigl[ \cos u \,{\bf e}_1 + \sin u \,{\bf e}_2 \Bigr] - \frac{\sinh\bigl(\tfrac{p+2}{2}v\bigr)}{p+2} \Bigl[ \cos\bigl(\tfrac{p+2}{2}u\bigr) {\bf e}_1 + \sin\bigl(\tfrac{p+2}{2}u\bigr) {\bf e}_2 \Bigr] \\
& \quad - \frac{\sinh\bigl(\tfrac{p-2}{2}v\bigr)}{p-2} \Bigl[ \cos\bigl(\tfrac{p-2}{2}u\bigr) {\bf e}_1 - \sin\bigl(\tfrac{p-2}{2}u\bigr) {\bf e}_2 \Bigr] + \frac{\sinh\bigl(\tfrac{p}{2}v\bigr)}{p/2} \sin\bigl(\tfrac{p}{2}u\bigr) {\bf e}_3 ,
\end{split}
\end{equation}
where $p$ is an arbitrary integer. The surfaces have the topology of an annulus with $p$ half-twists and $p$-fold rotational symmetry about the $3$-direction; they are orientable if $p$ is even and non-orientable if $p$ is odd. The M\"{o}bius strip ($p=1$) is known to be unstable~\cite{goldstein12} and instabilities in higher order twisted strips have also been observed~\cite{machon16}, leading to the formation of topological solitons in the non-orientable case. Some examples are shown in Figure~\ref{fig:fig_twist}.

The most common choice of Bj\"orling data to represent the circular helicoids is~\cite{meeks07,mira06}
\begin{gather}
{\bf c} = \cos u \,{\bf e}_1 + \sin u \,{\bf e}_2 ,
\label{eq:twisted_helicoid_Bc} \\
{\bf n} = \sin(pu/2) \bigl( \cos u \,{\bf e}_1 + \sin u \,{\bf e}_2 \bigr) + \cos(pu/2) {\bf e}_3 .
\label{eq:twisted_helicoid_Bn}
\end{gather}
For this Bj\"orling curve the dimensional reduction \eqref{eq:dim_red} leads to the operator 
\begin{equation}
- \partial_{uu} - 2 \Bigl[ (p/2)^2 + \sin^2(pu/2) \Bigr] ,
\end{equation}
whose eigenfunctions are Mathieu functions~\cite{McLachlan}. When $p$ is even the lowest eigenfunction is 
\begin{equation}
\phi(u) = \textrm{ce}_0\bigl( pu/2,2/p^2 \bigr) ,
\end{equation}
and has $p$-fold rotational symmetry, the same as the circular helicoid surface, while when $p$ is odd the lowest eigenfunction is $p$-fold degenerate and given by 
\begin{equation}
\phi(u) = \textrm{se}_{1/p}\bigl(p(u-u_0)/2,2/p^2\bigr) , 
\end{equation}
where $u_0=2\pi k/p$ with $k\in\mathbb{Z}/p\mathbb{Z}$ specifying one of $p$ equivalent locations for the single node in the eigenfunction. Using these eigenfunctions for $\phi$ and setting $\theta=0$ gives a set of Jacobi fields for the twisted strip, for which we were not able to produce exact expressions, however the eigenfunctions are well approximated by the first few terms in their Fourier expansion~\cite{McLachlan}. Numerical evidence suggests that the Jacobi fields produced in this way do not have bounded nodal domains for $p<6$, the first two examples with bounded domains are shown in Figure~\ref{fig:fig_twist}. 

Interestingly this set of Jacobi fields fails to reproduce an example of the unstable mode associated to the M\"{o}bius strip~\cite{goldstein12}. The choice of Jacobi field with $\theta=0$ ensures that the Bj\"{o}rling curve is a local maximum of $\psi$ in the transverse direction. Of course, there is a choice of Bj\"{o}rling curve for which a $\phi \neq 0$, $\theta=0$ Jacobi field reproduces the observed instabilities. We conclude, therefore, that an instability on the M\"{o}bius strip cannot be maximised on the circle~\eqref{eq:twisted_helicoid_Bc}, and so an alternate curve must be chosen. 

Nevertheless, our construction is, in general, capable of producing an arbitrary Jacobi field, and we demonstrate how choosing both $\phi$ and $\theta$ non-zero can reproduce an instability on the M\"{o}bius strip. We note that this provides an easy, analytically tractable alternative to the numerical solutions found previously~\cite{pesci15}. We make the following choices
\begin{equation}
\phi = \sin (u/2), \qquad \theta = \frac{1}{2} \sin u .
\end{equation}
This reproduces the support Jacobi field 
\begin{equation}
\psi = \frac{3+\cosh v - \cosh 2v + 8 \sinh^3(v/2) \cos(u/2)}{3 \cosh(v/2) \cosh v - 3 \cos(u/2) \sinh v} \, \sin(u/2) ,
\end{equation}
obtained by considering the derivative of the scaled family of surfaces ${\bf X}_t(u,v) = t {\bf X}(u,v)$ projected onto ${\bf N}$ at $t=1$. The critical domain of the M\"{o}bius strip found in this way is shown in Figure~\ref{fig:fig_twist}~(left).

%

\section{Discussion}

We have attempted to give a general description of instabilities in minimal surfaces and soap films by showing how variation of Bj\"orling data can be used to generate Jacobi fields. The approach is strongly geometric allowing for direct connection between the construction and experimental methodology. It can be developed for any curve on the minimal surface, including systolic curves where known which have been conjectured to be relevant to soap film collapses~\cite{goldstein14,pesci15,moffatt16}, although we have not proved any results towards choosing curves that are optimal in any sense and this remains an interesting outstanding problem. The approach also introduces a dimensional reduction of the stability operator for minimal surfaces to a one-dimensional Schr\"odinger operator. 
The examples we described subsume previous results on soap film instabilities and provide an interesting new description of the family of planar Enneper minimal surfaces. Many more explicit examples could be developed by utilising the methods of~\cite{lopez18} and an interesting extension of our work would be to study instabilities of periodic minimal surfaces. 

\vspace{5mm}
\acknowledgments{
We would like to thank J.H.~Hannay, R.B.~Kusner and G.~Rowlands for helpful discussions. G.P.A. supported in part by the UK EPSRC through Grant No. EP/N007883/1.
}

\end{document}